\documentclass[aps,prl,superscriptaddress,twocolumn,twoside,floatfix]{revtex4}

\usepackage{bm}
\usepackage{graphicx}
\usepackage{amsfonts}
\usepackage{amsmath}
\usepackage{amssymb}
\newcommand {\be}{\begin{equation}}
\newcommand {\ee}{\end{equation}}
\newcommand {\bea}{\begin{eqnarray}}
\newcommand {\eea}{\end{eqnarray}}

\begin{document}

\title{Strategies for optimal design for electrostatic energy storage in quantum multiwell heterostructures}
\author{Ilya Grigorenko}
\affiliation{Theoretical Division T-1, Center for Nonlinear Studies,
Center for Integrated Nanotechnologies, Los Alamos National
Laboratory, Los Alamos, New Mexico 87545, USA}
\author{Herschel Rabitz}
\affiliation{Chemistry Department, Princeton University, NJ 08544,
USA}
\date{\today}

\begin{abstract}
The physical principles are studied for the optimal design of a
quantum multiwell heterostructure working as an electrostatic energy
storage device. We performed the search for an optimal multiwell
trapping potential for electrons that results in the maximum static
palarizability of the system. The response of the heterostructure is
modeled quantum mechanically using nonlocal linear response theory.
Three main design strategies are identified, which lead to the
maximization of the stored energy. We found that the efficiency of
each strategy crucially depends on the temperature and the
broadening of electron levels. The energy density for optimized
heterostructures can exceed the nonoptimized value by a factor more
than $400$. These findings provide a basis for the development of
new nanoscale capacitors with high energy density storage
capabilities.
\end{abstract}

\maketitle

Attaining high density electromagnetic energy storage is a vital
technological and scientific problem, since electricity is the most
universal and scalable form of energy. In addition, the continued
progress in chip design  also produces demands for nanoscale
capacitors with the enhanced energy density. Recently, different
types of nanostructures, such as multi-walled doped carbon nanotubes
\cite{nisoli} or metal-dielectric interfaces \cite{interface,nature}
have attracted much attention as potential building blocks for high
energy density storage devices.

 There is no clear understanding of the physical principles
for optimal design on scales where quantum mechanical effects (e.
g., like discreteness of energy levels, tunneling  and nonlocality
of dielectric response) start to play a significant role.
 A recent study
\cite{NJP} predicted the surprising effect that there is an optimal
size for small metallic dimers resulting in maximal electromagnetic
energy density between the particles. This behavior is contrary to
the classical theory produces a monotonous increase of the energy
density with decreasing size of a dimer. Similar simulations, but
using more sophisticated {\it ab initio} techniques \cite{prl}
confirmed the correctness of qualitative predictions made in
\cite{NJP}. This confirmation provides a foundation for performing a
systematic study of the principles of optimal design of
nanocapacitors using the formalism developed \cite{prl2006}, thereby
avoiding much more computationally expensive {\it ab initio }
simulations.

The present study focuses on the static limit  of the dielectric
response of heterostructures. A dynamic field may result in higher
energy densities due to the resonance phenomena, however, this
benefit is offset by much higher losses in the dielectric, which is
not acceptable for energy storage applications. In the static
regime, there are still energy losses due to electron-phonon
coupling, etc. Typical values for the dissipation factor (reciprocal
to the quality $Q$ factor) is of the order $10^{-3}$ or less. In the
present study we assume zero dissipation factor in the static limit. 


This Letter considers a quantum multi-well heterostructure working
as an electrostatic energy storage device. We assume that the
heterostructure is placed between two leads, without direct contact
to the leads, permitting the neglect of tunneling effects. If a
constant bias is applied to the leads, the electron density in the
multi-well heterostructure becomes perturbed resulting in a finite
polarization of the nanostructure. The energy stored in the
heterostructure depends on the dielectric function of the
heterostructure, which is determined by the eigenfunctions and
eigenenergies of the confined electrons. The goal of optimal design
is to find a shape for the heterostructure, which maximizes the
stored electrostatic energy. In the present simulations a multi-well
heterostructure is modeled by the Hamiltonian $H=-{\bf
\nabla}^2+V_{trap}({\bf r})$, where $V_{trap}({\bf r})$ is the
electron effective trapping potential.

\begin{figure}
\includegraphics[width=4.2cm,angle=0]{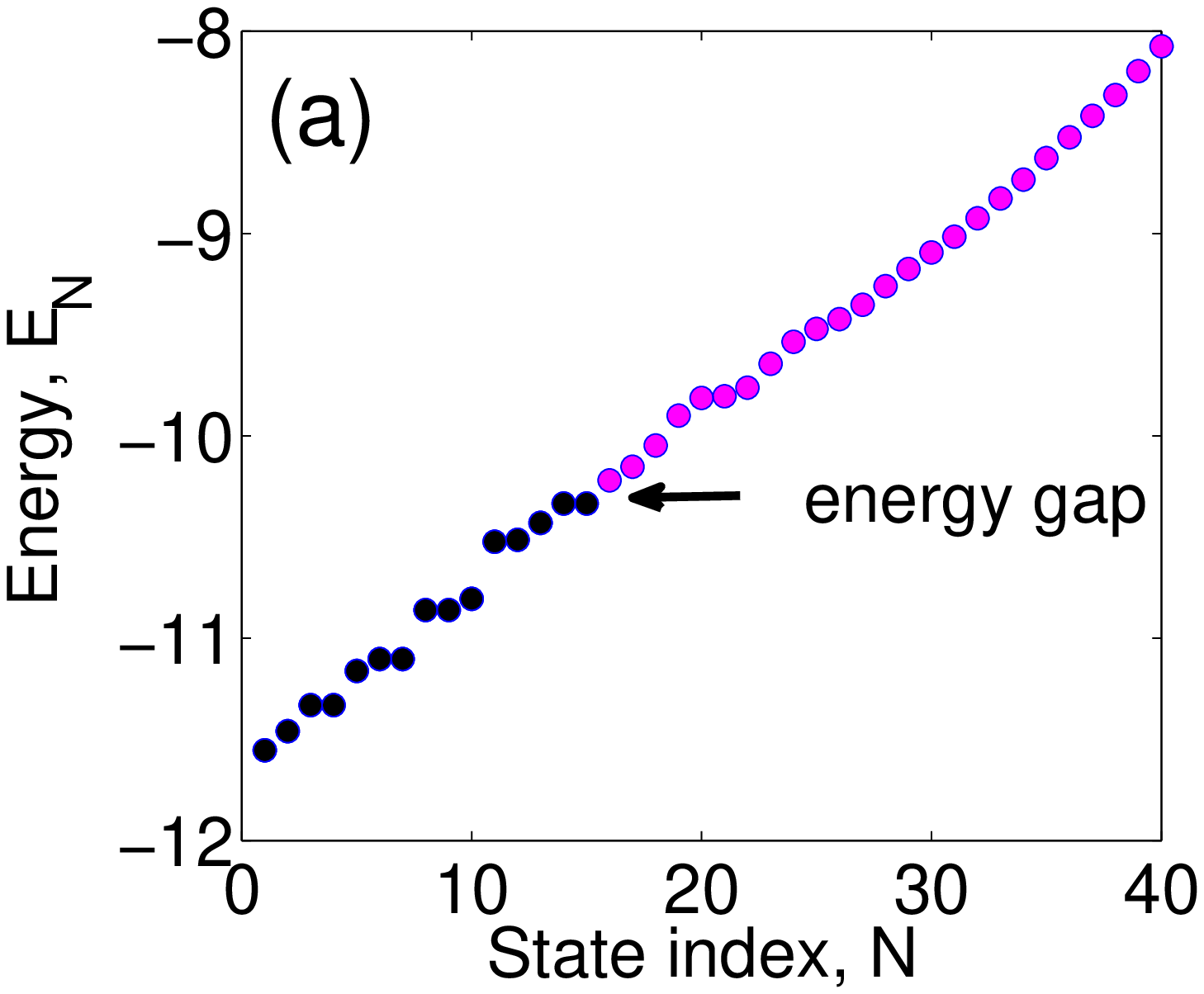}
\includegraphics[width=4.2cm,angle=0]{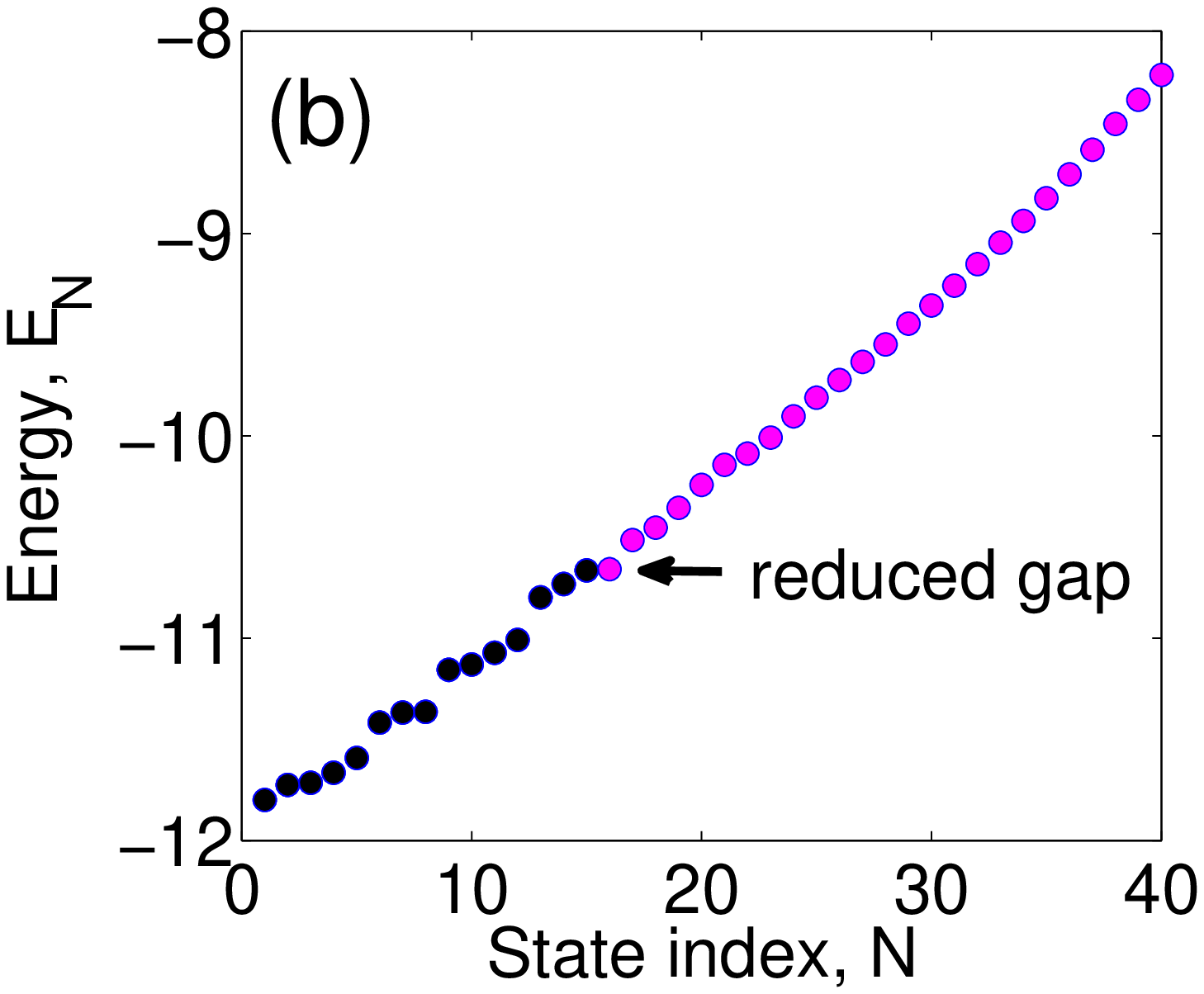}
\includegraphics[width=4.2cm,angle=0]{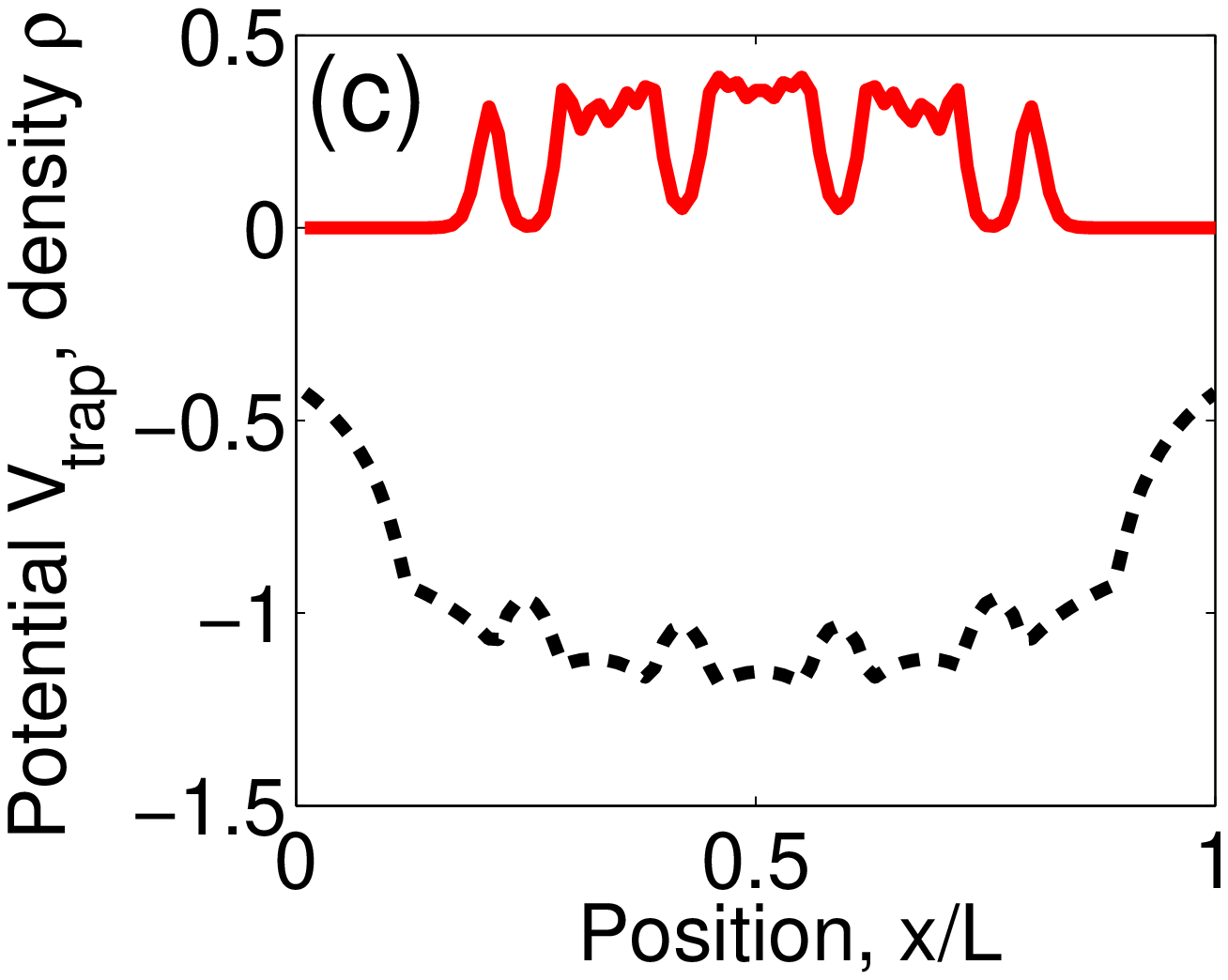}
\includegraphics[width=4.2cm,angle=0]{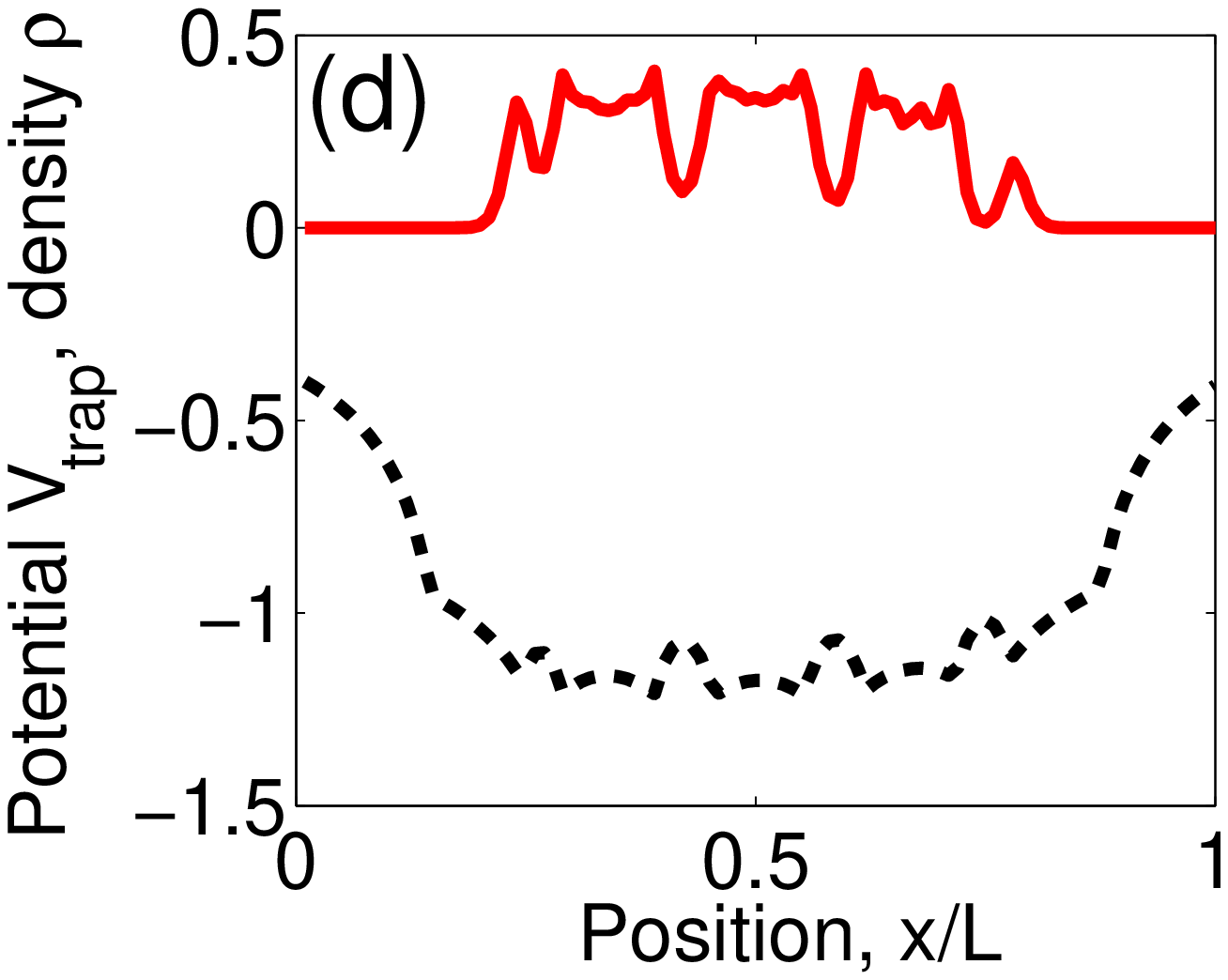}
\includegraphics[width=4.2cm,angle=0]{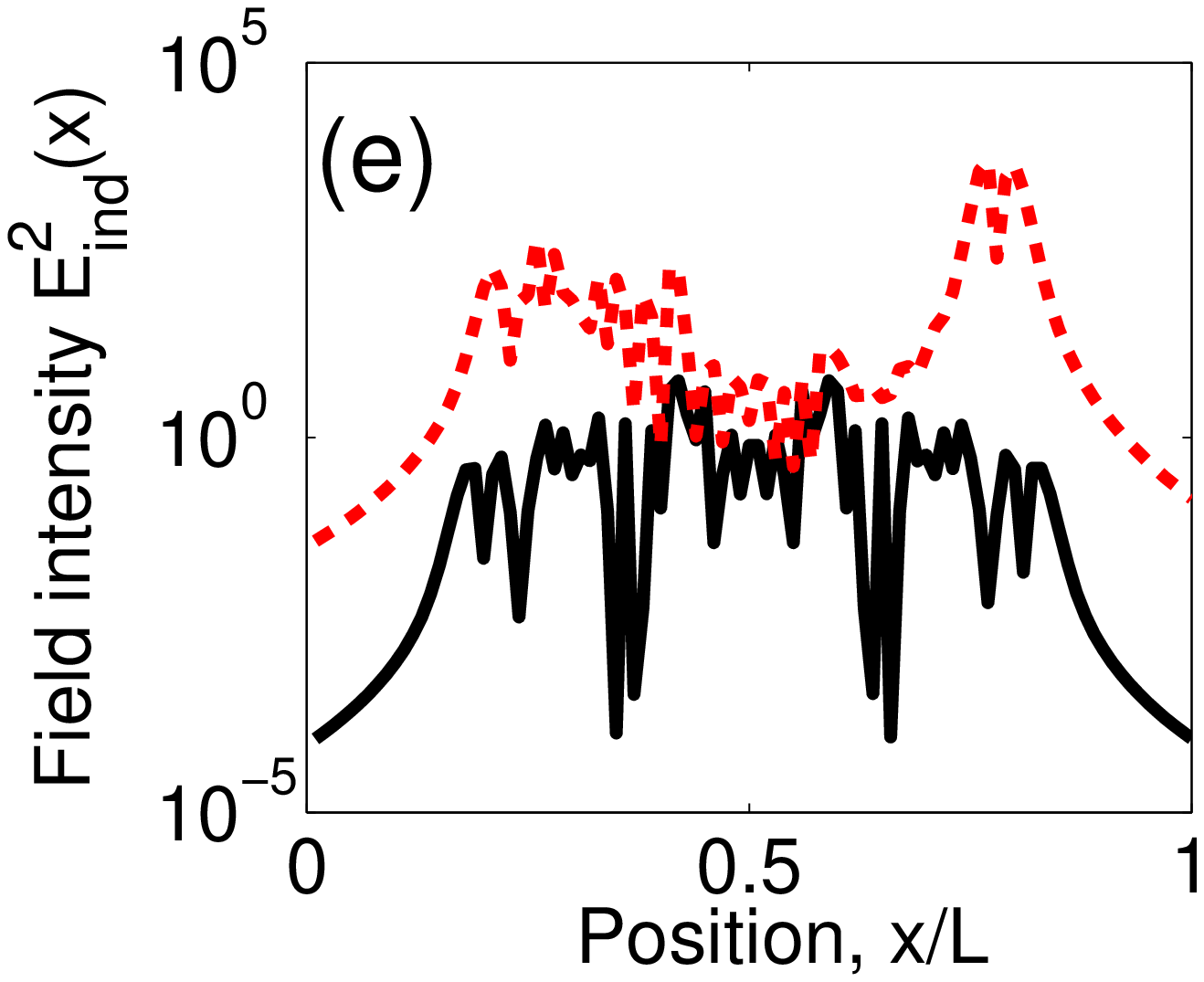}
\includegraphics[width=4.2cm,angle=0]{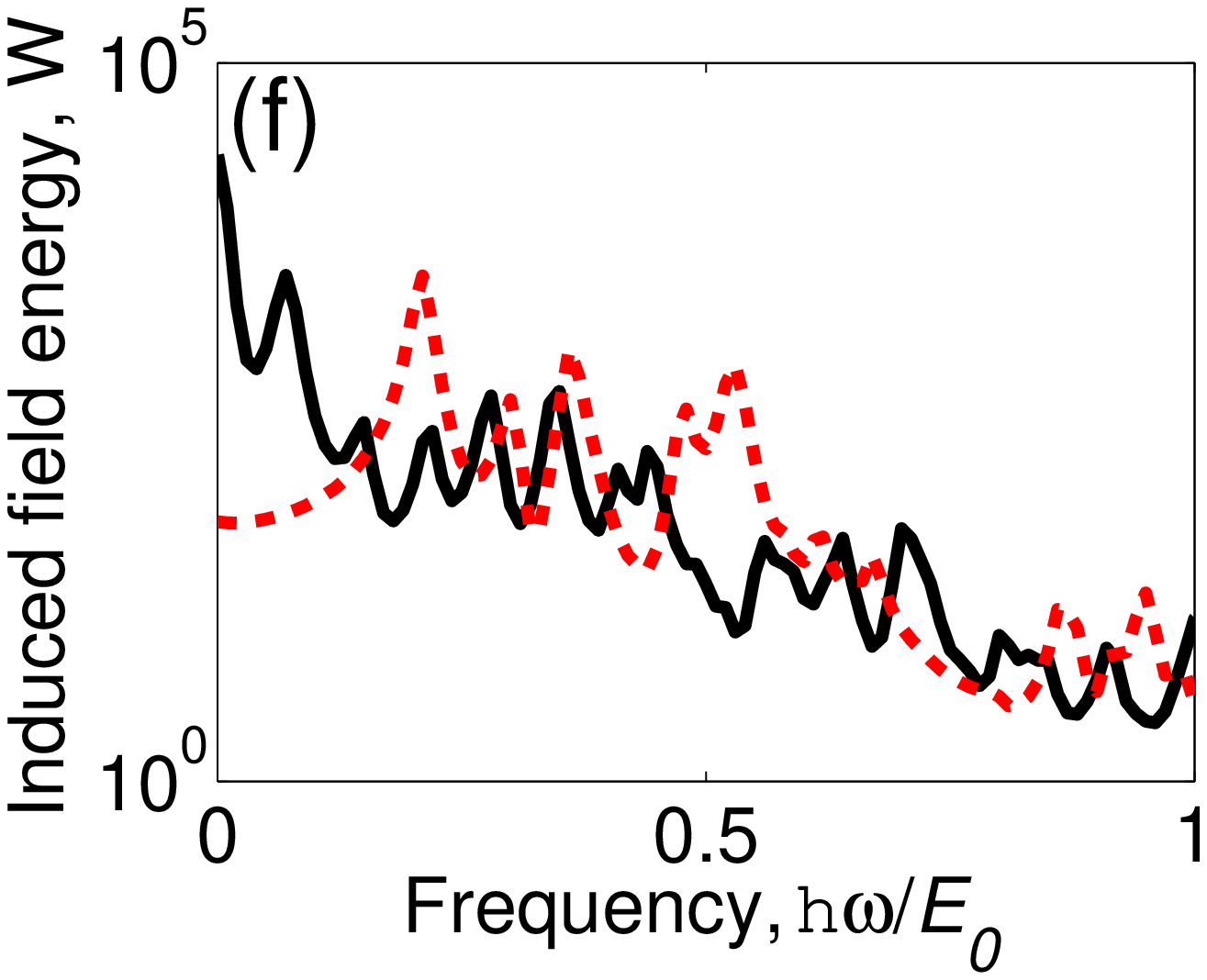}
\caption{\label{fig1} The eigenspectrum for a non-optimized
potential (a) and for the optimized one (b). The occupied states
colored in blue and unoccupied states are shown in magenta. Note,
that the energy gap $\Delta E_{N,N+1}$ in (b) is much smaller than
in (a). (c) The re-scaled trapping potential (black dashed line) and
the corresponding ground state electron density (red solid line) for
non-optimized  and optimized (d) configurations. (e) The induced
field intensity in the optimized trapping potential (red dashed
line) is enhanced by several orders of magnitude, compared to the
non-optimized potential (black solid line). (f) Frequency dependence
of the induced field energy in the nanostructure. Note, $T=0$,
$\gamma=0.01 E_0$.}
\end{figure}


 Since inhomogeneous systems with a few electrons can have an electromagnetic response very different from the bulk,
in our simulations we use the non-local density-density response
function within the coordinate-space representation $\chi ({\bf
r},{\bf r'},\omega )$ :
 \bea \label{chi}\chi ({\bf
r},{\bf r'},\omega ) = \sum\limits_{i,j} \frac{f(E_i ) - f(E_j
)}{E_i - E_j  + \hbar \omega  + i\gamma } \times \nonumber\\ \psi
_i^* ({\bf r}) \psi _i ({\bf r'})\psi _j^* ({\bf r'})\psi _j ({\bf
r}), \eea
 where $f(E_i)$ is the Fermi filling factor and the small constant
$\gamma$ describes the level broadening,
 and $E_i$ and $\psi_i$ are the eigenenergies and eigenfunctions of the Hamiltonian
 $H$, and
  $\omega$ is the frequency of
the external bias field. The electron eigenenergies $E_i$ and
eigenfunctions $\psi_i$ are obtained numerically using spacial
discretization of the Schr\"odinger equation.  Eq.(\ref{chi})
permits calculation of the induced electric field ${\bf
E}_{ind}({\bf r},\omega)$ in the system within linear response
theory \cite{prl2006}.

The effective trapping potential $V_{trap}({\bf r})$ is represented
as a sum of $N_p$ model potentials $V_{mod}({\bf r})$ located at the
positions ${\bf r}_i$, $i=1,..,N_p$, which may be controlled during
the material fabrication procedure: $V_{trap}({\bf
r})=\sum_{i=1}^{N_p} V_{mod}({\bf r}-{\bf r}_i)$. For simplicity  we
assume $V_{mod}({\bf r})$ to be the Coulomb potential of a charge
$Q$  with a cutoff $a$: $V_{mod}({\bf r})=Q/|{\bf r}|$, for $|{\bf
r}|>a$, $V_{mod}({\bf r})=Q/a$,  for $|{\bf r}|\le a$.
 The algorithm  searches for optimal parameters ${\bf r}_i$ to
maximize the stored electrostatic energy in the system $W=\int_V
|{\bf E}_{ind}({\bf r},\omega\to0)|^2dV$, where the integration is
performed over the volume of the nanostructure. The search for the
optimal potential $V_{trap}({\bf r})$ is performed using Brent's
``principal axis'' optimization algorithm \cite{brent} for a scalar
function of several variables.

To simplify our simulations we consider a quasi-one dimensional
system of length $L$. Then $E_0=\frac{\hbar^2}{2 m_e L^2}$ is the
unit of energy determined by
 the system, where $m_e$ is the electron mass. The
optimization of the trapping potential is performed at the static
limit of the external field $\hbar \omega=0$, with $N=15$ electrons
trapped in the heterostructure having $N_p=5$ wells. The cut-off
parameter is chosen as $a=0.05 L$, and $Q=3|e|$ to ensure the total
neutrality of the heterostructure. Note, that in all the simulations
we assume $\psi_i(0)=\psi(L)=0$, that is equivalent
$V_{trap}(0)=V_{trap}(L)=+\infty$. Fig.\ref{fig1}(a) shows the
eigenspectrum for a non-optimized potential, corresponding to
equally spaced potential wells (Fig.\ref{fig1}(c)). The occupied
states colored in blue and unoccupied states are shown in magenta.
Note, that the trapping potential is scaled down by the factor of
$10$ for better visibility.

First we perform the optimization for relatively small level
broadening $\gamma=0.01 E_0$ and temperature $T=0$K. The optimized
heterostructure has the eigenspectrum shown in Fig.\ref{fig1}(b),
and Fig.\ref{fig1}(d) shows the optimized trapping potential and the
corresponding ground state electron density. Note, that the
optimized potential has a dramatically decreased  energy gap $\Delta
E_{N,N+1}$ between the highest occupied and the lowest unoccupied
energy levels. The initial energy gap is $\Delta E_{N,N+1}\approx
0.11 E_0$, and after the optimization $\Delta E_{N,N+1}\approx 0.005
E_0$. For the field enhancement estimate we only need the  term in
Eq.(\ref{chi}) describing the transition between levels $N$ and
$N+1$. The decrease of the energy gap leads to an enhancement of the
$N\to N+1$ transition by approximately $\frac{|0.11-
0.01\:i|}{|0.005-0.01\:i|}\approx 10$, and the corresponding field
intensity (and stored energy $W$) by a factor of $\approx 100$,
compared to the non-optimized potential Fig.\ref{fig1}(e).
Potentially even further enhancement of the polarization may be
achieved, if there are several unoccupied energy levels (i.e. a
higher density of states), close to the highest occupied one.

 Since the energy $W$ stored in the heterostructure is increased by approximately factor of $450$
in the static limit (see Fig.\ref{fig1}(f)), it is clear, that the
vanishing of the gap is a significant, but not the only mechanism
for the field enhancement. The enhancement of the transition dipole
matrix elements, including for $N\to N+1$ is another mechanism.
However, for narrow energy levels this mechanism gives a relatively
less significant contribution of a factor of $\approx4$.

\begin{figure}
\includegraphics[width=4.2cm,angle=0]{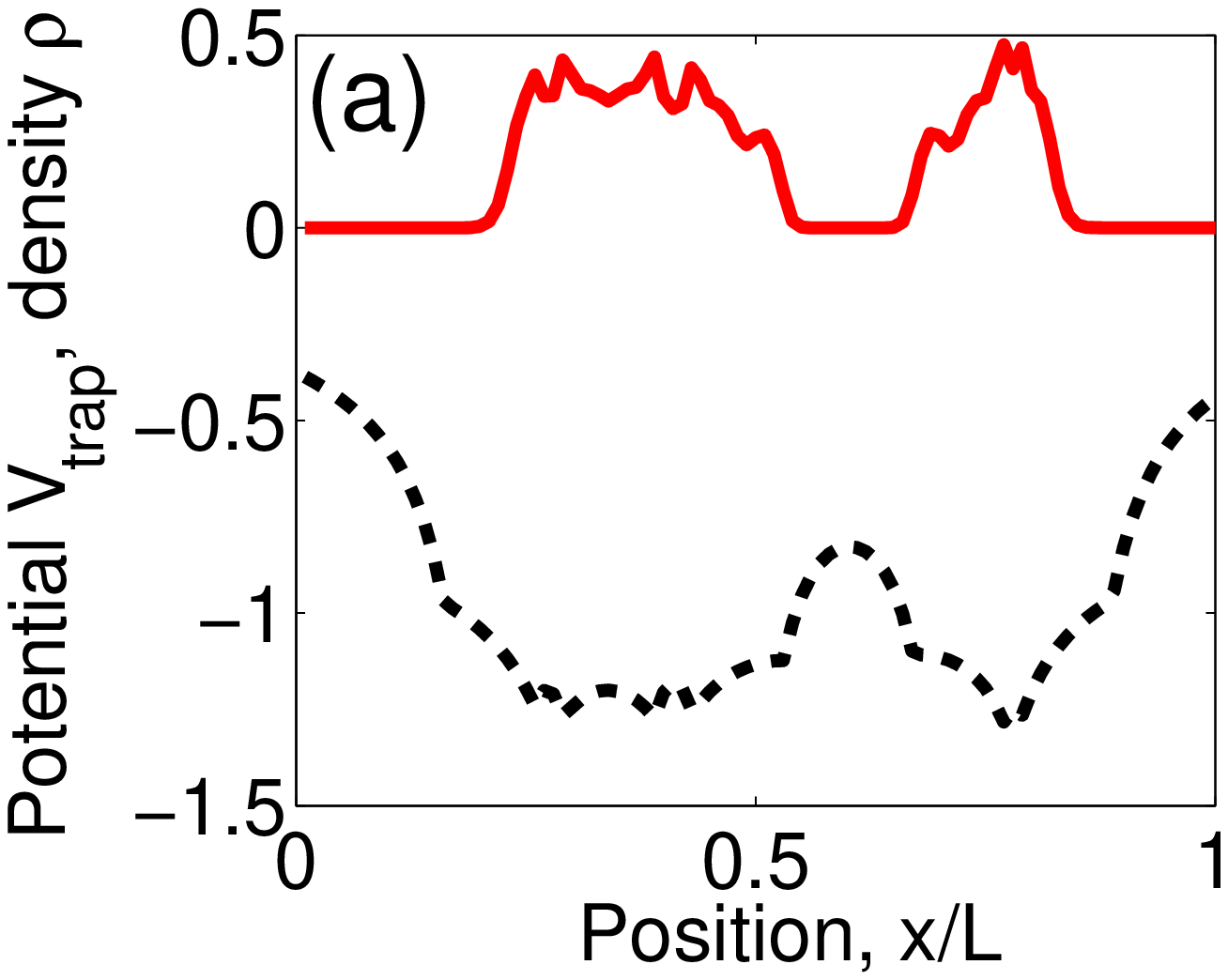}
\includegraphics[width=4.2cm,angle=0]{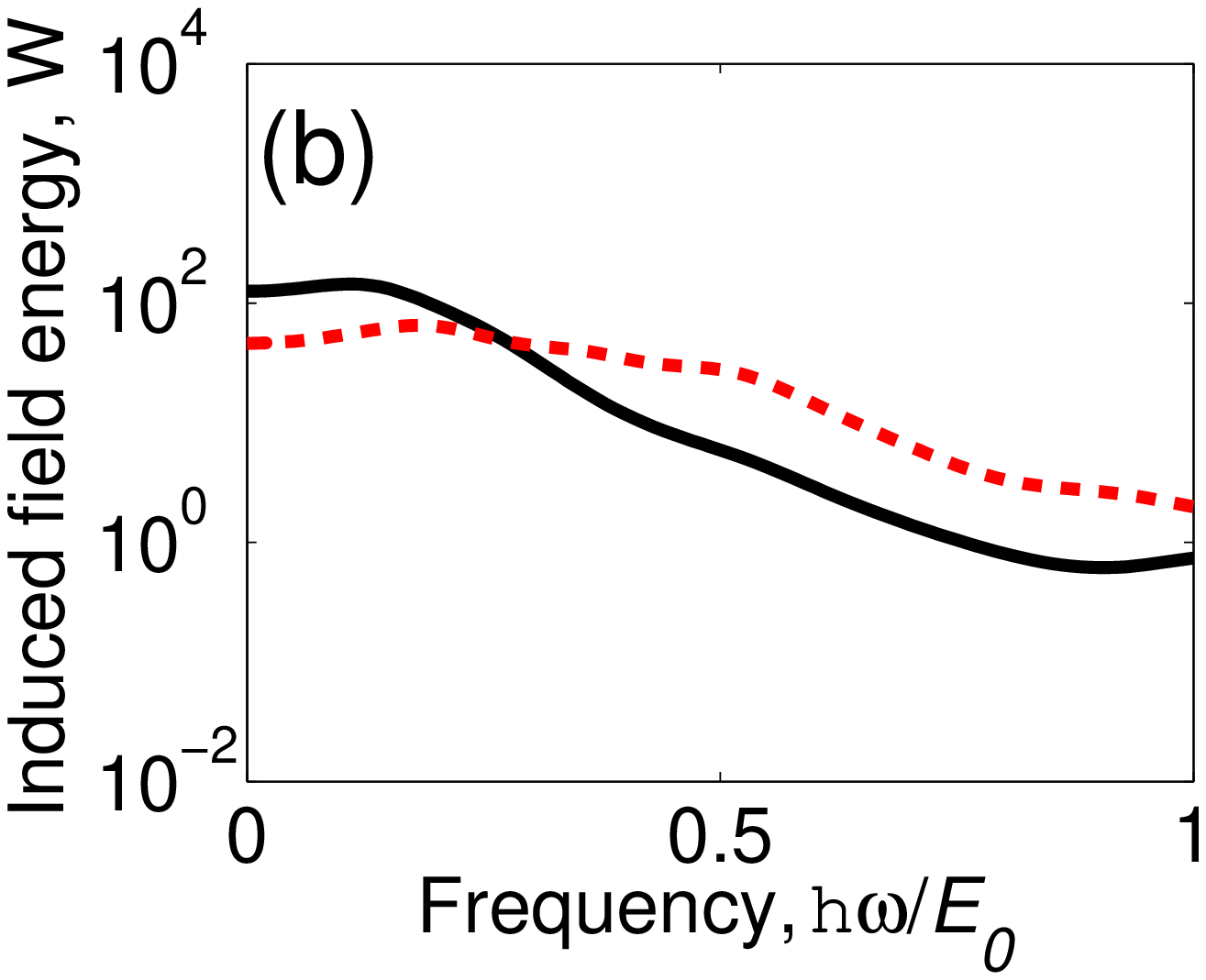}
\caption{\label{fig2}   (a) The effective trapping potential (black
dashed line) and the corresponding ground state electron density
(red solid line) for the optimized  heterostructure.  (b) Frequency
dependence of the induced field energy in the nanostructure. Note,
$T=0$, $\gamma=0.1 E_0$. }
\end{figure}

Upon increasing  the level broadening the dipole mechanism becomes
more significant. In Fig.\ref{fig2} we consider optimization of the
trapping potential in a case of larger level broadening $\gamma=0.1
E_0$.
 First, for the optimized
heterostructure in Fig.\ref{fig2}(a) the electron density  has the
shape with two maxima, resulting in a bigger dipole response of the
system. Second, for the optimized trapping potential the enhancement
of the energy stored in the heterostructure Fig.\ref{fig2}(b) is
much more modest, compared to Fig.\ref{fig1}(f). The optimized
design provides improvement of the stored energy by a factor
of $2.71$. Third, the energy gap $\Delta E_{N,N+1}\approx 0.02 E_0$
for the optimized potential is smaller, than in the non-optimized
one (see Fig.\ref{fig1}(a)), but it is still $\approx 4$ times
larger than in Fig.\ref{fig1}(b). Thus, the contribution due to the
resonant effect and the contribution due to the enhancement of the
dipole matrix elements are much closer, than in the previous
example, with narrower energy levels.
 Note, that the optimized  potential for $\gamma=0.01 E_0$
 produces significantly less polarization than
the optimized potential in Fig.\ref{fig2}(a). We conclude, that
 level broadening  makes the previously optimized design of the heterostructure non-optimal.

\begin{figure}
\includegraphics[width=4.2cm,angle=0]{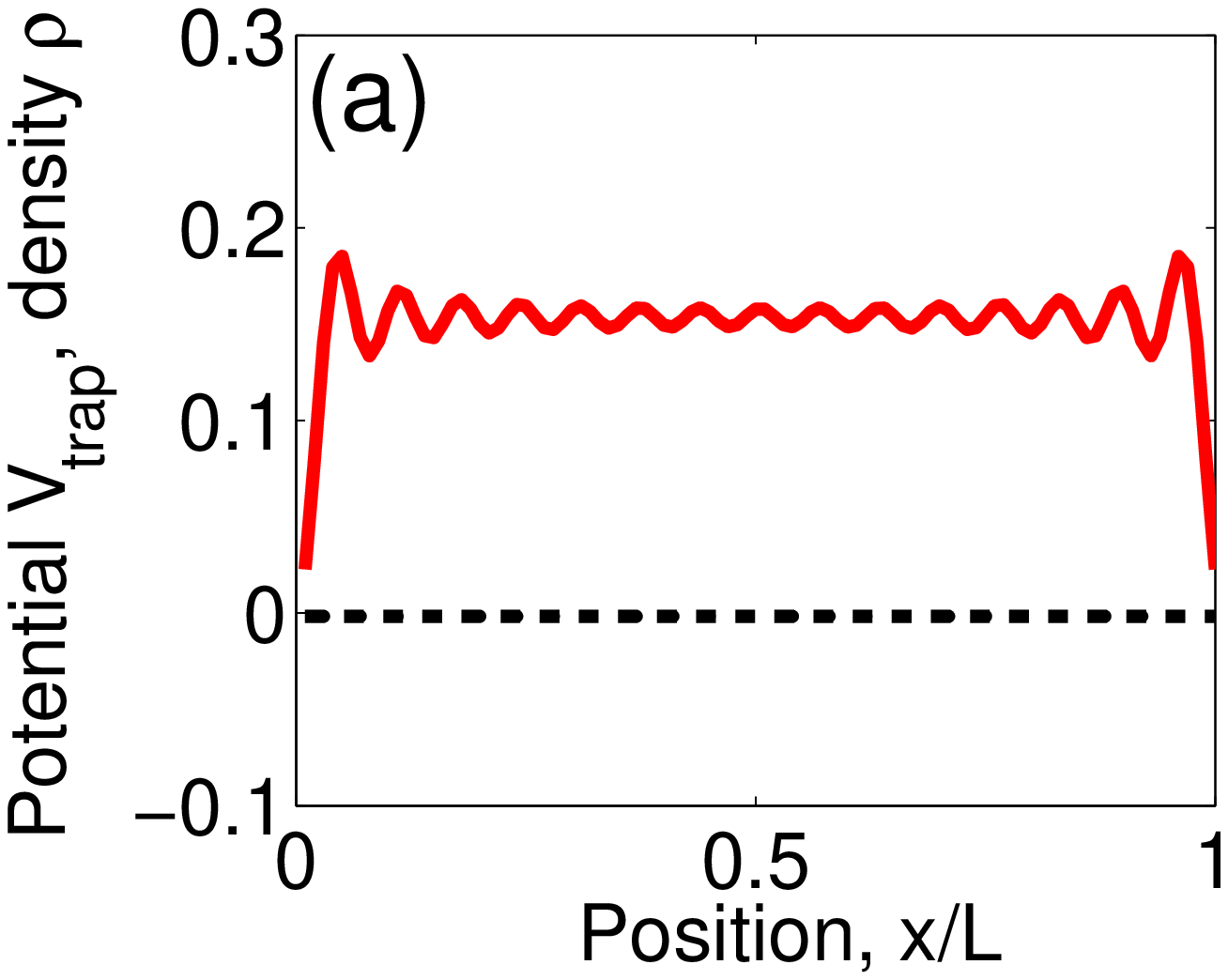}
\includegraphics[width=4.2cm,angle=0]{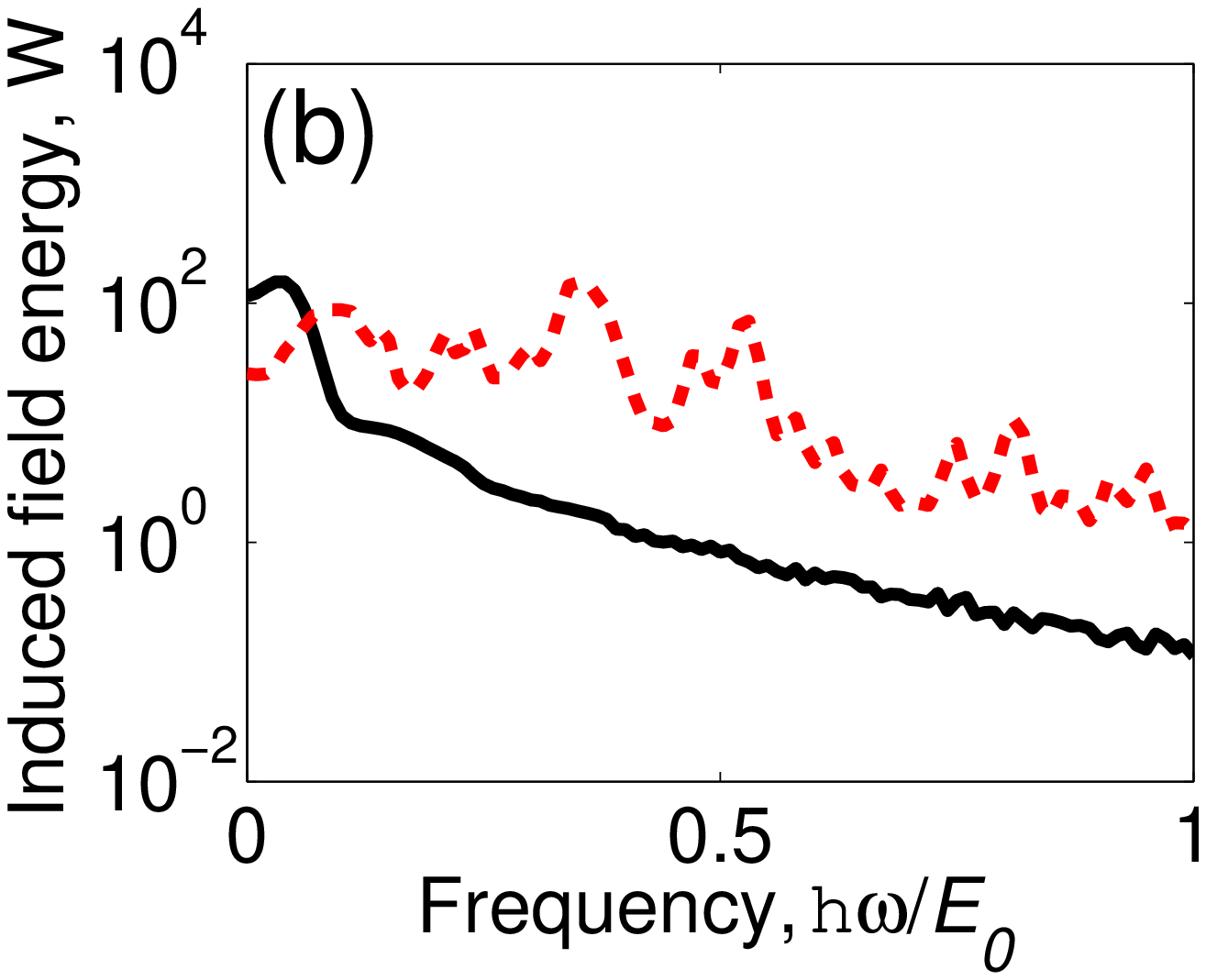}
\caption{\label{fig31} (a)  The effective trapping potential (black
dashed line) and the corresponding ground state electron density
(red solid line) for the optimized  configuration. (b) Frequency
dependence of the induced field energy in the nanostructure. Note,
$k_B T=0.5 E_0$, $\gamma=0.01 E_0$.}
\end{figure}

We also studied the optimization of the trapping potential at $T>0$K
in Fig.\ref{fig31}. Surprisingly, at the relatively high temperature
$k_B T=0.5E_0$ (here $k_B$ is the Boltzmann constant), the
optimization gives $V_{trap}(x)\equiv0$ as the best result,
corresponding to a square well potential. A reasonable explanation
of this surprising result is that the square well potential gives
the fastest scaling of the energy levels $E_N\propto N^2$, producing
the {\it maximum} energy gap $\Delta E_{N,N+1}$, and as a result,
the maximum population difference $f(E_{N+1})-f(E_{N})$.

In conclusion, we have studied the principles of optimal design of
heterostructures, which function as nanoscale capacitors. Three main
strategies to maximize the stored electrostatic energy were
identified. The first one is the minimization of the  energy gap
$\Delta E_{N,N+1}$, the second is the maximization of the transition
dipole matrix elements for the electrons in the heterostructure, and
the third  is the maximization of the population difference
$f(E_{N+1})-f(E_{N})$ in the system.  The effectiveness of each
strategy depends the level broadening and temperature of the system,
which we treated as independent parameters. Note, that the first and
the third strategies are mutually exclusive.

Some caution is necessary, as the minimization the energy gap
between the highest occupied and lowest unoccupied levels is
equivalent to making system ``more metallic''. In this case the
conductivity in such system should grow in the same way as the
susceptibility as a function of the energy gap $(\Delta
E_{N,N+1})^{-2}$, which is consistent with experimental findings
\cite{interface}. Since for energy storage applications one needs
nanocapacitors with relatively small inner charge leakage, the
scaling of the electrical conductivity and tunneling effects have to
be taken into account. Another effect, which we can not take into
account within linear response theory is the dielectric breakdown.
The increase of the breakdown voltage is favorable for energy
storage applications, since the electrostatic stored energy
$W\propto V_{max}^2$, where the maximum work voltage $V_{max}$
cannot exceed the breakdown limit. In general,  smaller energy gap
will result in a smaller critical breakdown voltage for bulk
materials, however, for spatially inhomogeneous nanostructures more
systematic {\it ab initio} simulations are desirable.

   \end{document}